# Mechanizing Set Theory

*Cardinal Arithmetic and the Axiom of Choice*

Lawrence C. Paulson
*Computer Laboratory, University of Cambridge*
*email: lcp@cl.cam.ac.uk*

Krzysztof Grabczewski
*Nicholas Copernicus University, Toruń, Poland*
*email: kgrabcze@mat.uni.torun.pl*

**Abstract.** Fairly deep results of Zermelo-Frænkel (ZF) set theory have been mechanized using the proof assistant Isabelle. The results concern cardinal arithmetic and the Axiom of Choice (AC). A key result about cardinal multiplication is $\kappa \otimes \kappa = \kappa$, where $\kappa$ is any infinite cardinal. Proving this result required developing theories of orders, order-isomorphisms, order types, ordinal arithmetic, cardinals, etc.; this covers most of Kunen, *Set Theory*, Chapter I. Furthermore, we have proved the equivalence of 7 formulations of the Well-ordering Theorem and 20 formulations of AC; this covers the first two chapters of Rubin and Rubin, *Equivalents of the Axiom of Choice*, and involves highly technical material. The definitions used in the proofs are largely faithful in style to the original mathematics.

# Contents

# 1. Introduction

A growing corpus of mathematics has been checked by machine. Researchers have constructed computer proofs of results in logic [26], number theory [25], group theory [28], $\lambda$-calculus [10], etc. An especially wide variety of results have been mechanized using the Mizar Proof Checker, including the theorem $\kappa \otimes \kappa = \kappa$ discussed below [2]. However, the problem of mechanizing mathematics is far from solved.

The Boyer/Moore Theorem Prover [3, 4] has yielded the most impressive results [25, 26]. It has been successful because of its exceptionally strong support for recursive definitions and inductive reasoning. But its lack of quantifiers forces mathematical statements to undergo serious contortions when they are formalized. Most automated reasoning systems are first-order at best, while mathematics makes heavy use of higher-order notations. We have conducted our work in Isabelle [20], which provides for higher-order syntax. Other recent systems that have been used for mechanizing mathematics include IMPS [6], HOL [8] and Coq [5].

We describe below machine proofs concerning cardinal arithmetic and the Axiom of Choice (AC). Paulson has mechanized most of the first chapter of Kunen [12] and a paper by Abrial and Laffitte [1]. Grąbczewski has mechanized the first two chapters of Rubin and Rubin's famous monograph [24], proving equivalent eight forms of the Well-ordering Theorem and twenty forms of AC. We have conducted these proofs using an implementation of Zermelo-Frænkel (ZF) set theory in Isabelle. Compared with other Isabelle/ZF proofs [15, 18, 21] and other automated set theory proofs [23], these are deep, abstract and highly technical results.

We have tried to reproduce the mathematics faithfully. This does not mean slavishly adhering to every detail of the text, but attempting to preserve its spirit. Mathematicians write in a mixture of natural language and symbols; they devise all manner of conventions to express their ideas succinctly. Their proofs make great intuitive leaps, whose detailed justification requires much additional work. We have been careful to note passages that seem unusually hard to mechanize, and discuss some of them below.

In conducting these proofs, particularly from Rubin and Rubin, we have tried to follow the footsteps of Jutting [11]. During the 1970s, Jutting mechanized a mathematics textbook using the AUTOMATH system [14]. He paid close attention to the text — which described the construction of the real and complex numbers starting from the Peano axioms — and listed any deviations from it. Compared with Jutting, we have worked in a more abstract field, and with source material containing larger gaps. But we have had the advantage of much more powerful hardware and software. We have relied upon Isabelle's reasoning tools (see §2 below) to fill some of the gaps for us.

We have done this work in the spirit of the QED Project [22], which aims "to build a computer system that effectively represents all important mathematical knowledge and techniques." Our results provide evidence, both positive and negative, regarding the feasibility of QED. On the positive side, we are able to mechanize difficult mathematics. On the negative side, the cost of doing so is hard to predict: a short passage can cause immense difficulties.

*Overview.* Section 2 is a brief introduction to Isabelle/ZF. The remaining sections report first Paulson's work and then Grąbczewski's. Sections 3–5 discuss the foundations of cardinal arithmetic in increasing detail, culminating in the machine proof of a key result about cardinal multiplication, $\kappa \otimes \kappa = \kappa$ where $\kappa$ is infinite. Section 6 introduces the Axiom of Choice and describes the mechanization of Abrial and Laffitte. Sections 7 and 8 are devoted to the mechanization of parts of Rubin and Rubin. Section 9 presents some conclusions.

## 2. Isabelle and ZF Set Theory

Isabelle [20] is a generic proof assistant. It supports proofs in higher-order logic, various modal logics, linear logic, etc. Our work is based upon Isabelle's implementation of Zermelo-Fraenkel (ZF) set theory, itself based upon an implementation of first-order logic. Isabelle/ZF arose from early work by Paulson [17] and Noël [15]; it is described in detail elsewhere [18, 21].

There are two key ideas behind Isabelle:

— *Expressions are typed $\lambda$-terms.* Thus the syntax is higher-order, giving a uniform treatment of quantifiers, descriptions and other binding operators. In Isabelle/ZF, all sets have the same type. But other important objects, such as classes, class relations and class functions, can be expressed using higher types.

— *Theorems are schematic inference rules.* Isabelle's basic inference mechanism is to join two schematic rules, in a sort of Horn clause resolution. A typical step in a backward proof consists of joining one rule (typically a lemma) to another rule (representing the proof state). Thus, theorems are proved by referring to previous theorems. Proof states may contain *unknowns*: placeholders for terms that have been left unspecified. Unification can incrementally instantiate unknowns, which may be shared among several subgoals.

Built around these key ideas are various facilities intended to ease the user's task. Notations can be defined using a general mixfix format, with precedences; variable-binding operators are easily specified. Isabelle manages a database of theories and theorems; when asked to load a theory, it automatically loads any other theories that it depends upon.

Although Isabelle supports proof checking, users will be more productive if they are provided with automatic tools.

- The *classical reasoner* solves subgoals using methods borrowed from tableau provers. It employs user-supplied rules, typically about logical connectives or set operators, to break down assertions.

- The *simplifier* employs user-supplied conditional equalities to rewrite a subgoal. It can make use of contextual information and handles commutative operators using a simple method borrowed from Boyer and Moore [3, page 104].

We have found these tools indispensable. But there is much room for improvement; mechanizing a page of text can take a week or more. We discuss some reasons for this below.

A lengthier introduction to Isabelle and Isabelle/ZF appears elsewhere [18]. The Isabelle documentation has been published as a book [20]. Figure 1 summarizes the Isabelle/ZF notation for set theory.

*Note.* Application of the function $f$ to the argument $x$ is formally written $f`x$. In informal mathematics we use the more familiar $f(x)$ for clarity. But a set-theoretic function is just another set, and Isabelle allows the notation $f(x)$ only if $f$ is a meta-level function. This usually corresponds to subscripting in informal mathematics, for example $f_x$. For the Isabelle/ZF development of functions, see Paulson [18, §7.5].

### 3. The Cardinal Proofs: Motivation and Discussion

The original reason for mechanizing the theory of cardinals was to generalize Paulson's treatment of recursive data structures in ZF. The original treatment [21] permitted only finite branching, as in $n$-ary trees. Countable branching required defining an uncountable ordinal. We are thus led to consider branching of any cardinality.

3.1. INFINITE BRANCHING TREES

Let $\kappa$ stand for an infinite cardinal and $\kappa^+$ for its successor cardinal. Branching by an arbitrarily large index set $I$ requires proving the theorem

$$\frac{|I| \leq \kappa \quad \forall_{i \in I} \alpha_i < \kappa^+}{(\bigcup_{i \in I} \alpha_i) < \kappa^+} \tag{1}$$

| syntax | description |
|---|---|
| $\{a_1, \ldots, a_n\}$ | finite set |
| `<a, b>` | ordered pair |
| $\{x\!:\!A \,.\, P[x]\}$ | Separation |
| $\{y \,.\, x\!:\!A, \; Q[x,y]\}$ | Replacement |
| $\{b[x] \,.\, x\!:\!A\}$ | functional Replacement |
| `INT` $x\!:\!A \,.\, B[x]$ | $\bigcap_{x \in A} . B[x]$, general intersection |
| `UN` $x\!:\!A \,.\, B[x]$ | $\bigcup_{x \in A} . B[x]$, general union |
| $A$ `Int` $B$ | $A \cap B$, intersection |
| $A$ `Un` $B$ | $A \cup B$, union |
| $A$ `->` $B$ | $A \to B$, function space |
| $A$ `*` $B$ | $A \times B$, Cartesian product |
| `PROD` $x\!:\!A \,.\, B[x]$ | $\Pi_{x \in A} \,.\, B[x]$, general product |
| `SUM` $x\!:\!A \,.\, B[x]$ | $\Sigma_{x \in A} \,.\, B[x]$, general sum |
| `THE` $x \,.\, P[x]$ | $\iota x \,.\, P[x]$, definite description |
| `lam` $x\!:\!A \,.\, b[x]$ | $\lambda_{x \in A} \,.\, b[x]$, abstraction |
| $f$ `'` $x$ | $f\text{'}x$ or $f(x)$, function application |
| $a$ `:` $A$ | $a \in A$, membership |
| $A$ `<=` $B$ | $A \subseteq B$, subset relation |
| `ALL` $x\!:\!A \,.\, P[x]$ | $\forall_{x \in A} \,.\, P[x]$, bounded quantifier |
| `EX` $x\!:\!A \,.\, P[x]$ | $\exists_{x \in A} \,.\, P[x]$, bounded quantifier |

*Figure 1.* ASCII notation for ZF

You need not understand the details of how this is used in order to follow the paper.[1]

Not many set theory texts cover such material well. Elementary texts [9, 27] never get far enough, while advanced texts such as Kunen [12] race through it. But Kunen's rapid treatment is nonetheless clear, and mentions all the essential elements. The desired result (1) follows fairly easily from Kunen's Lemma 10.21 [12, page 30]:

$$\frac{\forall_{\alpha < \kappa} |X_\alpha| \leq \kappa}{|\bigcup_{\alpha < \kappa} X_\alpha| \leq \kappa}$$

This, in turn, relies on the Axiom of Choice and its consequence the Well-ordering Theorem, which are discussed at length below. It also relies on a fundamental result about multiplication of infinite cardinals:

$$\kappa \otimes \kappa = \kappa.$$

This is Theorem 10.12 of Kunen. (In this paper, we refer only to his Chapter I.) The proof presents a challenging example of formalization, as we shall see.

We could prove $A \times A \approx A$, for all infinite sets $A$, by appealing to AC in the form of Zorn's Lemma; see Halmos [9, pages 97–8]. Then $\kappa \otimes \kappa = \kappa$ would follow immediately. But we need to prove $\kappa \otimes \kappa = \kappa$ without AC in order to use it in later proofs about equivalences of AC. In fact, the law $A \times A \approx A$ is known to be equivalent to the Axiom of Choice.

Paulson hoped to prove $\kappa \otimes \kappa = \kappa$ directly, but could not find a suitable proof. He therefore decided to mechanize the whole of Kunen's Chapter I, up to that theorem. We suggest this as a principle: theorems do not exist in isolation, but are part of a framework of supporting theorems. It is easier in the long run to build the entire framework, not just the parts thought to be relevant. The latter approach requires frequent, ad-hoc extensions to the framework.

3.2. OVERVIEW OF KUNEN, CHAPTER I

Kunen's first chapter is entitled, "Foundations of Set Theory." Kunen remarks on page 1 that the chapter is merely a review for a reader who has already studied basic set theory. This explains why the chapter is so succinct, as compared say with Halmos [9].

The first four sections are largely expository. Section 5 introduces a few axioms while §6 describes the operations of Cartesian product, relations, functions, domain and range. Already, §6 goes beyond the large Isabelle/ZF theory described in earlier papers [18, 21]. That theory emphasized computational notions, such as recursive data structures, at the expense of traditional set theory. Now it was time to develop some of the missing material. Paulson introduced some definitions about relations, orderings, well-orderings and order-isomorphisms, and proved the first two lemmas by well-founded induction. The main theorem required a surprising amount of further work; see §4.3 below.

Kunen's §7 covers ordinals. Much of this material had already been formalized in Isabelle/ZF [21, §3.2], but using a different definition of ordinal. A set $A$ is *transitive* if $A \subseteq \mathcal{P}(A)$: every element of $A$ is a subset of $A$. Kunen defines an ordinal to be a transitive set that is well-ordered by $\in$, while Isabelle/ZF defines an ordinal to be a transitive set of transitive sets. The two definitions are equivalent provided we assume, as we do, the Axiom of Foundation.

Our work required formalizing some material from §7 concerning order types and ordinal addition. We have also formalized ordinal multiplication. But we have ignored what Kunen calls $A^{<\omega}$ because Isabelle/ZF provides $\mathrm{list}(A)$, the set of finite lists over $A$ [21, §4.3] for the same purpose.

Kunen's §8 and §13 address the legitimacy of introducing new notations in axiomatic set theory. His discussion is more precise and comprehensive than Paulson's defence of the notation of Isabelle/ZF [18, page 361].

Kunen's §9 concerns classes and recursion. The main theorems of this section, justifying transfinite induction and recursion over the class of ordinals, were already in the Isabelle/ZF library [21, §3.2,§3.4]. Kunen discusses here (and with some irony in §12) the difficulties of formalizing properties of classes. Variables in ZF range over only sets; classes are essentially predicates, so a theorem about classes must be formalized as a theorem scheme.

Many statements about classes are easily expressed in Isabelle/ZF. An ordinary class is a unary predicate, in Isabelle/ZF an object of type $i \Rightarrow o$, where $i$ is the type of sets and $o$ is the type of truth values. A class relation is a binary predicate and has the Isabelle type $i \Rightarrow (i \Rightarrow o)$. A class function is traditionally represented by its graph, a single-valued class predicate [12, page 25]; it is more easily formalized in Isabelle as a meta-level function, an object of type $i \Rightarrow i$. See Paulson [18, §6] for an example involving the Replacement Axiom.

Because Isabelle/ZF is built upon first-order logic, quantification over variables of types $i \Rightarrow o$, $i \Rightarrow i$, etc., is forbidden. (And it should be; allowing such quantification in uses of the Replacement Axiom would be illegitimate.) However, schematic definitions and theorems may contain free variables of such types. Isabelle/ZF's transfinite recursion operator [21, §3.4] satisfies an equation similar to Kunen's Theorem 9.3, expressed in terms of class functions.

Isabelle/ZF does not overload set operators such as $\cap$, $\cup$, domain and list to apply to classes. Overloading is possible in Isabelle, but is probably not worth the trouble in this case. And the class-oriented definitions might be cumbersome. Serious reasoning about classes might be easier in some other axiomatic framework, where classes formally exist.

Kunen's §10 concerns cardinals. Some of these results presented great difficulties and form one of the main subjects of this paper. But the Schröder-Bernstein Theorem was already formalized in Isabelle/ZF [21, §2.6], and the first few lemmas were straightforward.

An embarrassment was proving that the natural numbers are cardinals. This boils down to showing that if there is a bijection between an $m$-element set and an $n$-element set then $m = n$. Proving this obvious fact is most tiresome. Reasoning about bijections is complicated; a helpful simplification (due to M. P. Fourman) is to reason about injections instead. Prove that if there is an injection from an $m$-element set to an $n$-element set then $m \leq n$. Applying this implication twice yields the desired result.

Many intuitively obvious facts are hard to justify formally. This came up repeatedly in our proofs, and slowed our progress considerably. It is a fundamental obstacle that will probably not yield to improved reasoning tools.

Kunen proves (Theorem 10.16) that for every ordinal $\alpha$ there is a larger cardinal, $\kappa$. Under AC this is an easy consequence of Cantor's Theorem; without AC more work is required. Paulson slightly modified Kunen's construction, letting $\kappa$ be the union of the order types of all well-orderings of subsets of $\alpha$, and found a pleasingly short machine proof.

Our main concern, as mentioned above, is Kunen's proof of $\kappa \otimes \kappa = \kappa$. We shall examine the machine proof in great detail. The other theorems of Kunen's

§10 concern such matters as cardinal exponentiation and cofinality. We have not mechanized these, but the only obstacle to doing so is time.

The rest of Kunen's Chapter I is mainly discussion.

## 4. Foundations of Cardinal Arithmetic

Let us examine the cardinal proofs in detail. We begin by reviewing the necessary definitions and theorems. Then we look at the corresponding Isabelle/ZF theories leading up to the main result, $\kappa \otimes \kappa = \kappa$. Throughout we shall concentrate on unusual aspects of the formalization, since much of it is routine.

### 4.1. Well-orderings

A relation $<$ is *well-founded* over a set $A$ provided every non-empty subset of $A$ has a $<$-minimal element. (This implies that $<$ admits no infinite decreasing chains $\cdots < x_n < \cdots < x_2 < x_1$ within $A$.) If furthermore $\langle A, < \rangle$ is a linear ordering then we say that $<$ *well-orders* $A$.

A function $f$ is an *order-isomorphism* (or just an *isomorphism*) between two ordered sets $\langle A, < \rangle$ and $\langle A', <' \rangle$ if $f$ is a bijection between $A$ and $A'$ that preserves the orderings in both directions: $x < y$ if and only if $f(x) <' f(y)$ for all $x, y \in A$.

Write $\langle A, < \rangle \cong \langle A', <' \rangle$ if there exists an order-isomorphism between $\langle A, < \rangle$ and $\langle A', <' \rangle$.

If $\langle A, < \rangle$ is an ordered set and $x \in A$ then $\mathrm{pred}(A, x, <) \stackrel{\mathrm{def}}{=} \{y \in A \mid y < x\}$ is called the (proper) *initial segment* determined by $x$. We also speak of $A$ itself as an initial segment of $\langle A, < \rangle$.

Kunen develops the theory of relations in his §6 and proves three fundamental properties of well-orderings:

- There can be no isomorphism between a well-ordered set and a proper initial segment of itself. A useful corollary is that if two initial segments are isomorphic to each other, then they are equal.

- There can be at most one isomorphism between two well-ordered sets. This result sounds important, but we have never used it.[2]

- Any two well-orderings are either isomorphic to each other, or else one of them is isomorphic to a proper initial segment of the other.

Kunen's proof of the last property consists of a single sentence:
Let $f =$

$$\{\langle v, w \rangle \mid v \in A \land w \in B \land \langle \mathrm{pred}(A, v, <_A) \rangle \cong \langle \mathrm{pred}(B, w, <_B) \rangle\};$$

note that $f$ is an isomorphism from some initial segment of $A$ onto some initial segment of $B$, and that these initial segments cannot both be proper.

This gives the central idea concisely; Suppes [27, pages 233–4] gives a much longer proof that is arguably less clear. However, the assertions Kunen makes are not trivial and Paulson needed two days and a half to mechanize the proof.

## 4.2. Order Types

The ordinals may be viewed as representatives of the well-ordered sets. Every ordinal is well-ordered by the membership relation $\in$. What is more important, every well-ordered set is isomorphic to a unique ordinal, called its *order type* and written $\mathrm{type}(A, <)$. Kunen [12, page 17] proves this by a direct construction. But to mechanize the result in Isabelle/ZF, it is easier to use well-founded recursion [21, §3.4]. If $\langle A, < \rangle$ is a well-ordering, define a function $f$ on $A$ by the recursion

$$f(x) = \{f(y) \mid y < x\}$$

for all $x \in A$. Then

$$\mathrm{type}(A, <) \stackrel{\mathrm{def}}{=} \{f(x) \mid x \in A\}.$$

It is straightforward to show that $f$ is an isomorphism between $\langle A, < \rangle$ and $\mathrm{type}(A, <)$, which is indeed an ordinal.

Our work has required proving many properties of order types, such as methods for calculating them in particular cases. Our source material contains few such proofs; we have spent much time re-discovering them.

## 4.3. Combining Well-orderings

Let $A + B \stackrel{\mathrm{def}}{=} (\{0\} \times A) \cup (\{1\} \times B)$ stand for the disjoint sum of $A$ and $B$, which is formalized in Isabelle/ZF [21, §4.1]. Let $\langle A, <_A \rangle$ and $\langle B, <_B \rangle$ be well-ordered sets. The order types of certain well-orderings of $A + B$ and $A \times B$ are of key importance.

The sum $A + B$ is well-ordered by a relation $<$ that combines $<_A$ and $<_B$, putting the elements of $A$ before those of $B$. It satisfies the following rules:

$$\frac{a' <_A a}{\mathrm{Inl}(a') < \mathrm{Inl}(a)} \qquad \frac{b' <_B b}{\mathrm{Inr}(b') < \mathrm{Inr}(b)} \qquad \frac{a \in A \quad b \in B}{\mathrm{Inl}(a) < \mathrm{Inr}(b)}$$

The product $A \times B$ is well-ordered by a relation $<$ that combines $<_A$ and $<_B$, lexicographically:

$$\frac{a' <_A a \quad b', b \in B}{\langle a', b' \rangle < \langle a, b \rangle} \qquad \frac{a \in A \quad b' <_B b}{\langle a, b' \rangle < \langle a, b \rangle}$$

```
Cardinal = OrderType + Fixedpt + Nat + Sum +
consts
  Least            :: (i=>o) => i     (binder "LEAST " 10)
  eqpoll, lepoll,
         lesspoll :: [i,i] => o       (infixl 50)
  cardinal         :: i=>i            ("|_|")
  Finite, Card    :: i=>o

defs
  Least_def      "Least(P) == THE i. Ord(i) & P(i) &
                                     (ALL j. j<i --> ~P(j))"
  eqpoll_def     "A eqpoll B == EX f. f: bij(A,B)"
  lepoll_def     "A lepoll B == EX f. f: inj(A,B)"
  lesspoll_def   "A lesspoll B == A lepoll B & ~(A eqpoll B)"
  Finite_def     "Finite(A) == EX n:nat. A eqpoll n"
  cardinal_def   "|A| == LEAST i. i eqpoll A"
  Card_def       "Card(i) == (i = |i|)"
end
```

*Figure 2.* Isabelle/ZF Theory Defining the Cardinal Numbers

The well-orderings of $A + B$ and $A \times B$ are traditionally used to define the ordinal sum and product. We do not require ordinal arithmetic until we come to the proofs from Rubin and Rubin. But we require the well-orderings themselves in order to prove $\kappa \otimes \kappa = \kappa$. That proof requires yet another well-ordering construction: *inverse image*.

If $\langle B, <_B \rangle$ is an ordered set and $f$ is a function from $A$ to $B$ then define $<_A$ by

$$x <_A y \iff f(x) <_B f(y).$$

Clearly $<_A$ is well-founded if $<_B$ is. If $f$ is injective and $<_B$ is a well-ordering then $<_A$ is also a well-ordering. If $f$ is bijective then obviously $f$ is an isomorphism between the orders $\langle A, <_A \rangle$ and $\langle B, <_B \rangle$; it follows that their order types are equal.

Sum, product and inverse image are useful for expressing well-orderings; this follows Paulson's earlier work [16] within Constructive Type Theory.

4.4. CARDINAL NUMBERS

Figure 2 presents the Isabelle/ZF definitions of cardinal numbers, following Kunen's §10. The Isabelle theory file extends some Isabelle theories (Order-Type and others) with constants, which stand for operators or predicates. The constants are defined essentially as follows:

− The least ordinal $\alpha$ such that $P(\alpha)$ is defined by a unique description [18, pages 366–7] and may be written LEAST $\alpha$ . $P(\alpha)$.

- Two sets $A$ and $B$ are *equipollent* if there exists a bijection between them. Write $A \approx B$ or, in Isabelle, $A$ `eqpoll` $B$.

- $B$ *dominates* $A$ if there exists an injection from $A$ into $B$. Write $A \precsim B$ or $A$ `lepoll` $B$.

- $B$ *strictly dominates* $A$ if $A \precsim B$ and $A \not\approx B$. Write $A \prec B$ or $A$ `lesspoll` $B$.

- A set is *finite* if it is equipollent to a natural number.

- The *cardinality* of $A$, written $|A|$, is the least ordinal equipollent to $A$. Without AC, no such ordinal has to exist; we might then regard $|A|$ as undefined. But everything is defined in Isabelle/ZF. The operator `THE` returns 0 unless the description identifies an object uniquely. Thus, an "undefined" cardinality equals 0; this conveniently ensures that $|A|$ is always an ordinal.

- A set $i$ is a *cardinal* if $i = |i|$; write `Card`$(i)$.

Reasoning from these definitions is entirely straightforward except for the "obvious" facts about finite cardinals mentioned above.

### 4.5. Cardinal Arithmetic

Let $\kappa$, $\lambda$, $\mu$ range over finite or infinite cardinals. Cardinal sum and product are defined in terms of disjoint sum and Cartesian product:

$$\kappa \oplus \lambda \stackrel{\text{def}}{=} |\kappa + \lambda|$$
$$\kappa \otimes \lambda \stackrel{\text{def}}{=} |\kappa \times \lambda|$$

These satisfy the familiar commutative, associative and distributive laws. The proofs are uninteresting but non-trivial, especially as we work without AC. We do so in order to use the results in proving various forms of AC to be equivalent (see below); but frequently this forces us to construct well-orderings explicitly.

## 5. Proving $\kappa \otimes \kappa = \kappa$

We begin with an extended discussion of Kunen's proof and then examine its formalization.

### 5.1. Kunen's Proof

Kunen calls this result Theorem 10.12. His proof is admirably concise.

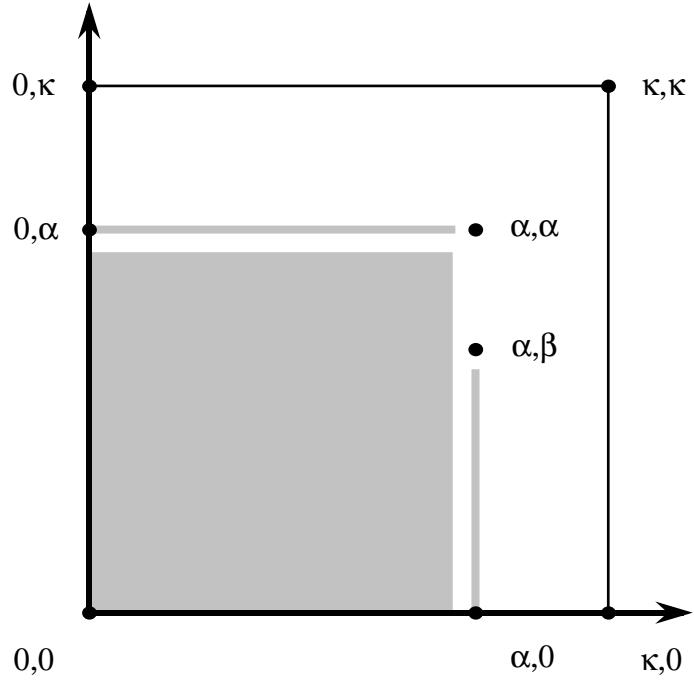

*Figure 3.* Predecessors of $\langle \alpha, \beta \rangle$, with $\beta \leq \alpha$

**Theorem.** If $\kappa$ is an infinite cardinal then $\kappa \otimes \kappa = \kappa$.

**Proof.** By transfinite induction on $\kappa$. Assume this holds for smaller cardinals. Then for $\alpha < \kappa$, $|\alpha \times \alpha| = |\alpha| \otimes |\alpha| < \kappa$ (applying Lemma 10.10 when $\alpha$ is finite).[3] Define a well-ordering $\triangleleft$ on $\kappa \times \kappa$ by $\langle \alpha, \beta \rangle \triangleleft \langle \gamma, \delta \rangle$ iff

$$\max(\alpha, \beta) < \max(\gamma, \delta) \ \lor \ [\max(\alpha, \beta) = \max(\gamma, \delta) \land \\ \langle \alpha, \beta \rangle \text{ precedes } \langle \gamma, \delta \rangle \text{ lexicographically}].$$

Each $\langle \alpha, \beta \rangle \in \kappa \times \kappa$ has no more than

$$|\mathrm{succ}(\max(\alpha, \beta)) \times \mathrm{succ}(\max(\alpha, \beta))| < \kappa$$

predecessors in $\triangleleft$, so $\mathrm{type}(\kappa \times \kappa, \triangleleft) \leq \kappa$, whence $|\kappa \times \kappa| \leq \kappa$. Since clearly $|\kappa \times \kappa| \geq \kappa$, $|\kappa \times \kappa| = \kappa$.

The key to the proof is the ordering $\triangleleft$, whose structure may be likened to that of a square onion. Let $\alpha$ and $\beta$ be ordinals such that $\beta \leq \alpha < \kappa$. The predecessors of $\langle \alpha, \beta \rangle$ include all pairs of the form $\langle \alpha, \beta' \rangle$ for $\beta' < \beta$, and all pairs of the form $\langle \alpha', \alpha \rangle$ for $\alpha' < \alpha$; these pairs constitute the $\alpha^{\text{th}}$ layer of the onion. The other predecessors of $\langle \alpha, \beta \rangle$ are pairs of the form $\langle \gamma, \delta \rangle$ such that $\gamma, \delta < \alpha$; these pairs constitute the inner layers of the onion. (See Figure 3.)

The set of all ◁-predecessors of $\langle\alpha,\beta\rangle$ is a subset of $\mathrm{succ}(\alpha)\times\mathrm{succ}(\alpha)$, which gives an upper bound on its cardinality. Kunen expresses this upper bound in terms of $\max(\alpha,\beta)$ to avoid having to assume $\beta\leq\alpha$.

To simplify the formal proofs, Paulson used the more generous upper bound

$$|\mathrm{succ}(\mathrm{succ}(\max(\alpha,\beta)))\times\mathrm{succ}(\mathrm{succ}(\max(\alpha,\beta))))|.$$

This is still a cardinal below $\kappa$. As Kunen notes, there are two cases. If $\alpha$ or $\beta$ is infinite then $\mathrm{succ}(\mathrm{succ}(\max(\alpha,\beta)))<\kappa$ because $\max(\alpha,\beta)<\kappa$ and because infinite cardinals are closed under successor; therefore, the inductive hypothesis realizes our claim. On the other hand, if $\alpha$ and $\beta$ are both finite, then so is $\mathrm{succ}(\mathrm{succ}(\max(\alpha,\beta)))$, while $\kappa$ is infinite by assumption.

To complete the proof, we must examine the second half of Kunen's sentence: "so $\mathrm{type}(\kappa\times\kappa,◁)\leq\kappa$, whence $|\kappa\times\kappa|\leq\kappa$." Recall from §4.2 that there is an isomorphism

$$f:\kappa\times\kappa\to\mathrm{type}(\kappa\times\kappa,◁)$$

such that

$$f(\alpha,\beta)=\{f(\gamma,\delta)\mid\langle\gamma,\delta\rangle◁\langle\alpha,\beta\rangle\}.$$

Thus, $f(\alpha,\beta)$ is an ordinal with the same cardinality as the set of predecessors of $\langle\alpha,\beta\rangle$. This implies $f(\alpha,\beta)<\kappa$ for all $\alpha,\beta<\kappa$, and therefore $\mathrm{type}(\kappa\times\kappa,◁)\leq\kappa$. Because $f$ is a bijection between $\kappa\times\kappa$ and $\mathrm{type}(\kappa\times\kappa,◁)$, we obtain $|\kappa\times\kappa|\leq\kappa$. The opposite inequality is trivial.

### 5.2. Mechanizing the Proof

Proving $\kappa\otimes\kappa=\kappa$ requires formalizing the relation ◁. Kunen's definition looks complicated, but we can get the same effect using our well-ordering constructors (recall §4.3). Note that ◁ is an inverse image of the lexicographic well-ordering of $\kappa\times\kappa\times\kappa$, under the function $g:\kappa\times\kappa\to\kappa\times\kappa\times\kappa$ defined by

$$g(\alpha,\beta)=\langle\max(\alpha,\beta),\alpha,\beta\rangle;$$

this function is trivially injective.

Figure 4 presents part of the Isabelle theory file for cardinal arithmetic. It defines ◁ as the constant `csquare_rel`. Here is a summary of the operators appearing in its definition:

- `rvimage`$(A,f,<)$ is the inverse image ordering on $A$ derived from $<$ by $f$.

- `lam <x,y>:K*K. <x Un y, x, y>` is the function called $g$ above. The pattern-matching in the abstraction expands internally to the constant `split`, which takes apart ordered pairs [18, page 367]. Finally `Un` denotes union; note that $\max(\alpha,\beta)=\alpha\cup\beta$ for ordinals $\alpha$ and $\beta$.

```
CardinalArith = Cardinal + OrderArith + Arith + Finite +
consts
  InfCard      :: i=>o
  "|*|"        :: [i,i]=>i       (infixl 70)
  "|+|"        :: [i,i]=>i       (infixl 65)
  csquare_rel  :: i=>i

defs
  InfCard_def  "InfCard(i) == Card(i) & nat le i"
  cadd_def     "i |+| j == |i+j|"
  cmult_def    "i |*| j == |i*j|"

  csquare_rel_def
  "csquare_rel(K) ==
   rvimage(K*K,
           lam <x,y>:K*K. <x Un y, x, y>,
           rmult(K, Memrel(K),
                 K*K, rmult(K,Memrel(K),K,Memrel(K))))"
end
```

*Figure 4.* Isabelle/ZF Theory File for Cardinal Arithmetic

- $\mathtt{rmult}(A, <_A, B, <_B)$ constructs the lexicographic ordering on $A \times B$ from the orderings $<_A$ and $<_B$.

- $\mathtt{Memrel}(\kappa)$ is the membership relation on $\kappa$. This is the primitive well-ordering for ordinals.

Proving that csquare_rel is a well-ordering is easy, thanks to lemmas about rvimage and rmult. A single command proves that our map is injective.

Figure 5 presents the nine theorems that make up the Isabelle/ZF proof of $\kappa \otimes \kappa = \kappa$. The theorems are stated literally in Isabelle notation. The symbol ==> expresses implication from premises to conclusion. Multiple premises are bracketed using [ | and | ]. For example, theorem 2 is the inference

$$\frac{\mathtt{Ord}(\kappa)}{\mathtt{well\_ord}(\kappa \times \kappa, \mathtt{csquare\_rel}(\kappa))}$$

and theorem 3 is

$$\frac{x < \kappa \quad y < \kappa \quad z < \kappa \quad \langle\langle x,y\rangle, \langle z,z\rangle\rangle \in \mathtt{csquare\_rel}(\kappa)}{x \leq z \wedge y \leq z}$$

There is not enough space to present the proofs, which comprise over sixty Isabelle tactic commands; see Paulson [18, §8] for demonstrations of tactics. The nine proofs require a total of 43 seconds to run.[4]

The first few theorems concern elementary properties of $\mathtt{csquare\_rel}(\kappa)$. We find that it is a well-ordering of $\kappa$ (theorems 1, 2) and that the initial segment below $\xi$, for $\xi < \kappa$, is a subset of $\mathrm{succ}(\xi) \times \mathrm{succ}(\xi)$ (theorems 3, 4). The

```
1  Ord(K) ==>
   (lam <x,y>:K*K. <x Un y, x, y>) : inj(K*K, K*K*K)

2  Ord(K) ==> well_ord(K*K, csquare_rel(K))

3  [| x<K;  y<K;  z<K;  <<x,y>, <z,z>> : csquare_rel(K) |] ==>
   x le z & y le z

4  z<K ==> pred(K*K, <z,z>, csquare_rel(K)) <= succ(z)*succ(z)

5  [| x<z;  y<z;  z<K |] ==> <<x,y>, <z,z>> : csquare_rel(K)

6  [| InfCard(K);  x<K;  y<K;  z=succ(x Un y) |] ==>
   ordermap(K*K, csquare_rel(K)) ` <x,y>  <
   ordermap(K*K, csquare_rel(K)) ` <z,z>

7  [| InfCard(K);  x<K;  y<K;  z=succ(x Un y) |] ==>
   |ordermap(K*K, csquare_rel(K)) ` <x,y>|   le
   |succ(z)|  |*|  |succ(z)|

8  [| InfCard(K);  ALL y:K. InfCard(y) --> y |*| y = y |]  ==>
   ordertype(K*K, csquare_rel(K)) le K

9  InfCard(K) ==> K |*| K = K
```

*Figure 5.* Theorems for the Proof of $\kappa \otimes \kappa = \kappa$

next three theorems (5, 6, 7) form part of the proof that $\kappa$ is the order type of csquare_rel($\kappa$). The isomorphism called $f$ in §5.1 is written in Isabelle/ZF as

    ordermap(K*K, csquare_rel(K)).

If $\alpha, \beta < \kappa$ then, setting $\xi = \mathrm{succ}(\mathrm{succ}(\max(\alpha,\beta)))$, we obtain $f(\alpha,\beta) \precsim f(\xi,\xi)$ and thus, via theorem 4, we have $|f(\alpha,\beta)| \leq |\xi| \otimes |\xi|$.

Theorem 7 corresponds to the first part of Kunen's sentence, "Each $\langle \alpha, \beta \rangle \in \kappa \times \kappa$ has no more than $|\mathrm{succ}(\max(\alpha,\beta)) \times \mathrm{succ}(\max(\alpha,\beta))|$ predecessors in $\triangleleft$," and it took about a day to prove. Theorem 8 covers the next part of the sentence, "so $\mathrm{type}(\kappa \times \kappa, \triangleleft) \leq \kappa$," and took another day to prove. This theorem assumes the transfinite induction hypothesis in order to verify $|\mathrm{succ}(\xi)| \otimes |\mathrm{succ}(\xi)| \leq \kappa$ in the case when $\xi$ is infinite, checking the finite case separately. At 17 tactic steps, the proof is the most complicated of the nine theorems. The main result, theorem 9, merely sets up the transfinite induction and appeals to the previous theorems.

Kunen uses without proof the analogous result for addition of infinite cardinals, $\kappa \oplus \kappa = \kappa$. We could prove it using an argument like the one above, but with an ordering of $\kappa + \kappa$ instead of $\kappa \times \kappa$. Fortunately there is a much simpler proof, combining the trivial $\kappa \leq \kappa \oplus \kappa$ with the chain of inequalities $\kappa \oplus \kappa = 2 \otimes \kappa \leq \kappa \otimes \kappa = \kappa$. Formalized mathematics requires discovering such simple proofs whenever possible.

The effort required to prove $\kappa \otimes \kappa = \kappa$ includes not only the several days spent formalizing the few sentences of Kunen's proof, but also the weeks spent developing a library of results about orders, well-orderings, isomorphisms, order types, cardinal numbers and basic cardinal arithmetic. After proving the theorem, more work was required to complete the theoretical foundation for infinite branching trees (recall our original motivation, §3.1). Fortunately, we have been able to re-use the libraries for proofs about AC. This we turn to next.

### 6. The Axiom of Choice and the Well-Ordering Theorem

Our construction of infinite branching trees uses the Axiom of Choice. Let us review the main features of this axiom and consider how to formalize it in Isabelle.

If $C$ is a set of non-empty sets then AC asserts that there is a function $f$ such that $f(c) \in c$ for all $c \in C$. We can formalize this straightforwardly as

$$\frac{0 \notin C}{\exists_{f \in C \to \bigcup(C)} \forall_{c \in C} \, f`c \in c}$$

Replacing the function space $C \to \bigcup(C)$ by a general product is less familiar but more concise and direct:

$$\frac{0 \notin C}{\exists f \,.\, f \in \prod_{c \in C} c}$$

We call $f \in (\prod_{c \in C} c)$ a *choice function* on the set $C$.

Expressing the set $C$ in different forms, such as $\mathcal{P}(A) - \{0\}$ or $\{B(x) \mid x \in A\}$, yields various equivalent assertions of AC. Isabelle/ZF follows Halmos [9] in expressing AC as *the product of a family of non-empty sets is non-empty*. It derives many equivalent formulations of AC. All this is done in a separate Isabelle theory of AC, which can be imported when necessary; most of Isabelle/ZF is developed without AC.

AC is significant only when applied to an infinite set. If $c \neq 0$ then, trivially, there exists $x \in c$. Let $C$ be a finite set. Then the other axioms of set theory let us construct a choice function by induction on the size of $C$. One can express weaker choice principles by restricting $C$.

By eliminating $C$ altogether we could obtain a stronger axiom, *Global Choice*:

$$\frac{c \neq 0}{\mathrm{choice}(c) \in c}$$

The choice operator is, in effect, a choice function on the universe itself. It is easy to use but formally stronger than we need.[5]

The Well-Ordering Theorem states that every set can be well-ordered. It implies that $|A|$ is always meaningful. Every set is equipollent to some ordinal, namely the order type of some well-ordering. The least such ordinal is its cardinality.

It is not hard to see that the Well-Ordering Theorem is equivalent to AC. If we apply the theorem to the set $\bigcup(C)$ then we can define a choice function $f$ such that, for $c \in C$, $f(c)$ yields the least element of $c$ under the well-ordering. Conversely, if we apply AC to the set $\mathcal{P}(A) - \{0\}$ then we can repeatedly choose new elements of $A$ to construct a well-ordering. The details are messy.

Kunen assumes AC in the form of the Well-Ordering Theorem, perhaps to avoid those messy details, but Isabelle/ZF tackles this proof. Fortunately, Abrial and Laffitte describe the proof with the aim of mechanization [1]. Starting from AC they prove Hausdorff's Maximal Principle, Zorn's Lemma and the Well-Ordering Theorem. Paulson mechanized their proofs easily. There are under 180 tactic commands, which take about 140 seconds to execute.

Abrial and Laffitte describe their research as a study about proofs. They work in a typed version of Zermelo set theory. The proofs hold in standard ZF set theory too, though as the authors remark, there are simpler proofs for ZF. This does not disturb us because their exposition saves us a great deal of effort.

Their proofs are more detailed than necessary even for mechanization. They devote a full page to Lemma 0, a result about unions; Isabelle's classical reasoner can prove this unaided in 1.4 seconds:

```
goal ZF.thy "!!A B C. (ALL x:C. x<=A | B<=x) ==>
                      Union(C)<=A | B<=Union(C)";
by (fast_tac ZF_cs 1);
```

Their proofs are based upon the original work of Zermelo. Instead of using the ordinals, they make an inductive definition similar to the construction of the ordinals but taking the successor operation as a parameter. Provided the successor operation satisfies certain conditions, the inductive set turns out to be totally ordered by inclusion ($\subseteq$), in fact well-ordered. Then, supplying suitable successor operations allows proving the desired results, such as the well-ordering theorem.

Mechanizing these proofs did present a few challenges. Their proof of the Well-Ordering Theorem appears to contain an error; we used an alternative justification of their Property 6.4. The inductive definition involves fixedpoints and some non-trivial proofs, but Isabelle's inductive definition package [19] automates this process. Abrial and Laffitte envisaged the definition and related proofs to depend implicitly on its successor parameter. In Isabelle this parameter must be explicit in all definitions and proofs, and its assumed properties must be stated wherever they are needed. This did not cause major complications, but it might have done so.

Abrial and Laffitte adopt the Axiom of Global Choice and use the choice operator in definitions. Since we do not have this operator, many of our theorems take the existence of a choice function as an additional assumption. When AC is finally invoked, the rule of existential elimination discharges this assumption.

Their formal language resembles higher-order logic. Their paper is thus relevant to many proof assistants, such as HOL [8], IMPS [6] and Isabelle/HOL [20]. We have used it to define Isabelle/ZF's library of the main forms of AC. But this hardly exhausts the subject. Rather, it is merely the introduction to our next case study.

## 7. Rubin and Rubin's AC Proofs

Herman and Jean Rubin's book *Equivalents of the Axiom of Choice* [24] is a compendium of hundreds of statements equivalent to the Axiom of Choice. Many of these statements were used originally as formulations of AC; others, of independent interest, were found to be equivalent to AC. Each chapter of the book focusses on a particular framework for formulating AC. Chapter 1 discusses equivalent forms of the Well-Ordering Theorem. Chapter 2 discusses the Axiom of Choice itself. Other chapters cover the Trichotomy Law, cardinality formulations, etc.

Grąbczewski has mechanized the first two chapters, both definitions and proofs. He has additionally proved the equivalence of all the formulations given; the book omits the "easy" proofs and a few of the harder ones. Below we outline the definitions and some of the more interesting proofs.

This is a substantial piece of work. There are 55 definitions, mostly names of the formulations of AC. There are nearly 1900 tactic commands. The full development takes over 44 minutes to run.[6]

### 7.1. THE WELL-ORDERING THEOREM

The eight equivalent forms of the Well-Ordering Theorem are the following:

$WO_1$ Every set can be well-ordered.

$WO_2$ Every set is equipollent to an ordinal number.

$WO_3$ Every set is equipollent to a subset of an ordinal number.

$WO_4(m)$ For every set $x$ there exists an ordinal $\alpha$ and a function $f$ defined on $\alpha$ such that $f(\beta) \precsim m$ for every $\beta < \alpha$ and $\bigcup_{\beta < \alpha} f(\beta) = x$.

$WO_5$ There exists a natural number $m \geq 1$ such that $WO_4(m)$.

```
WO1_def   "WO1 == ALL A. EX R. well_ord(A,R)"

WO2_def   "WO2 == ALL A. EX a. Ord(a) & A eqpoll a"

WO3_def   "WO3 == ALL A. EX a. Ord(a) & (EX b. b<=a & A eqpoll b)"

WO4_def   "WO4(m) == ALL A. EX a f. Ord(a) & domain(f)=a &
                     (UN b<a. f'b) = A & (ALL b<a. f'b lepoll m)"

WO5_def   "WO5 == EX m:nat. 1 le m & WO4(m)"

WO6_def   "WO6 == ALL A. EX m:nat. 1 le m & (EX a f. Ord(a) &
                  domain(f)=a & (UN b<a. f'b) = A &
                  (ALL b<a. f'b lepoll m))"

WO7_def   "WO7 == ALL A. Finite(A) <-> (ALL R. well_ord(A,R) -->
                  well_ord(A,converse(R)))"

WO8_def   "WO8 == ALL A. (EX f. f : (PROD X:A. X)) -->
                  (EX R. well_ord(A,R))"
```

*Figure 6.* Isabelle/ZF Definitions of Well-Ordering Principles

$\text{WO}_6$ For every set $x$ there exists a natural number $m \geq 1$, an ordinal $\alpha$, and a function $f$ defined on $\alpha$ such that $f(\beta) \precsim m$ for every $\beta < \alpha$ and $\bigcup_{\beta < \alpha} f(\beta) = x$.

$\text{WO}_7$ For every set $x$, $x$ is finite iff for each well-ordering $R$ of $x$, $R^{-1}$ also well-orders $x$.

$\text{WO}_8$ Every set possessing a choice function can be well-ordered.

Most of Chapter 1 is devoted to proving $\text{WO}_6 \Longrightarrow \text{WO}_1$, which is by far the hardest of the results. Grąbczewski has proved the equivalence of all the formulations given above by means of the following implications:

$$\begin{aligned}
\text{WO}_1 &\Longrightarrow \text{WO}_2 \Longrightarrow \text{WO}_3 \Longrightarrow \text{WO}_1 \\
&\text{WO}_4(m) \Longrightarrow \text{WO}_4(n) \quad \text{if } m \leq n \\
\text{WO}_4(n) \Longrightarrow \text{WO}_5 &\Longrightarrow \text{WO}_6 \Longrightarrow \text{WO}_1 \Longrightarrow \text{WO}_4(1) \\
\text{WO}_7 &\iff \text{WO}_1 \\
\text{WO}_8 &\iff \text{WO}_1
\end{aligned}$$

Figure 6 shows how these axioms are formalized in Isabelle.

### 7.2. THE AXIOM OF CHOICE

The formulations of the Axiom of Choice are as follows:

$AC_1$ If $A$ is a set of non-empty sets, then there is a function $f$ such that for every $B \in A$, $f(B) \in B$.

$AC_2$ If $A$ is a set of non-empty, pairwise disjoint sets, then there is a set $C$ whose intersection with any member $B$ of $A$ has exactly one element.

$AC_3$ For every function $f$ there is a function $g$ such that for every $x$, if $x \in \text{dom}(f)$ and $f(x) \neq 0$, then $g(x) \in f(x)$.

$AC_4$ For every relation $R$ there is a function $f \subseteq R$ such that $\text{dom}(f) = \text{dom}(R)$.

$AC_5$ For every function $f$ there is a function $g$ such that $\text{dom}(g) = \text{range}(f)$ and $f(g(x)) = x$ for every $x \in \text{dom}(g)$.

$AC_6$ The Cartesian product of a set of non-empty sets is non-empty.

$AC_7$ The Cartesian product of a set of non-empty sets of the same cardinality is non-empty.

$AC_8$ If $A$ is a set of pairs of equipollent sets, then there is a function which associates with each pair a bijection mapping one onto the other.

$AC_9$ If $A$ is a set of sets of the same cardinality, then there is a function which associates with each pair a bijection mapping one onto the other.

$AC_{10}(n)$ If $A$ is a set of sets of infinite sets, then there is a function $f$ such that for each $x \in A$, the set $f(x)$ is a decomposition of $x$ into disjoint sets of size between 2 and $n$.

$AC_{11}$ There exists a natural number $n \geq 2$ such that $AC_{10}(n)$.

$AC_{12}$ If $A$ is a set of sets of infinite sets, then there is a natural number $n \geq 2$ and a function $f$ such that for each $x \in A$, the set $f(x)$ is a decomposition of $x$ into disjoint sets of size between 2 and $n$.

$AC_{13}(m)$ If $A$ is a set of non-empty sets, then there is a function $f$ such that for each $x \in A$, the set $f(x)$ is a non-empty subset of $x$ with $f(x) \precsim m$.

$AC_{14}$ There is a natural number $m \geq 1$ such that $AC_{13}(m)$.

$AC_{15}$ If $A$ is a set of non-empty sets, then there is a natural number $m \geq 1$ and a function $f$ such that for each $x \in A$, the set $f(x)$ is a non-empty subset of $x$ with $f(x) \precsim m$.

$AC_{16}(n,k)$ If $A$ is an infinite set, then there is a set $t_n$ of $n$-element subsets of $A$ such that each $k$-element subset of $A$ is a subset of exactly one element of $t_n$.

$AC_{17}$ If $A$ is a set, $B = \mathcal{P}(A) - \{0\}$ and $g$ is a function from $B \to A$ to $B$, then there is a function $f \in B \to A$ such that $f(g(f)) \in g(f)$.

$AC_{18}$ For every non-empty set $A$, every family of non-empty sets $\{B_a \mid a \in A\}$ and every family of sets $\{X_{a,b} \mid a \in A, b \in B_a\}$, there holds[7]

$$\bigcap_{a \in A} \bigcup_{b \in B_a} X_{a,b} = \bigcup_{f \in \prod_{a \in A} B_a} \bigcap_{a \in A} X_{a,f(a)}.$$

$AC_{19}$ For any non-empty set $A$, each of whose elements is non-empty,

$$\bigcap_{a \in A} \bigcup_{b \in a} b = \bigcup_{f \in C(A)} \bigcap_{a \in A} f(a),$$

where $C(A)$ is the set of all choice functions on $A$.

Grąbczewski has mechanized the following proofs in Isabelle:

$$\begin{array}{cc}
AC_1 \iff AC_2 & AC_4 \iff AC_5 \\
AC_1 \iff AC_6 & AC_6 \iff AC_7 \\
\multicolumn{2}{c}{AC_1 \implies AC_4 \implies AC_3 \implies AC_1} \\
\multicolumn{2}{c}{AC_1 \implies AC_8 \implies AC_9 \implies AC_1} \\
\multicolumn{2}{c}{WO_1 \implies AC_1 \implies WO_2} \\
\multicolumn{2}{c}{WO_1 \implies AC_{10}(n) \implies AC_{11} \implies AC_{12} \implies AC_{15} \implies WO_6} \\
AC_{10}(n) \implies AC_{13}(n-1) & AC_{13}(n) \implies AC_{14} \implies AC_{15} \\
\multicolumn{2}{c}{AC_{11} \implies AC_{14}} \\
\multicolumn{2}{c}{AC_{13}(m) \implies AC_{13}(n) \quad \text{if } m \le n} \\
AC_1 \iff AC_{13}(1) & AC_1 \iff AC_{17} \\
\multicolumn{2}{c}{WO_2 \implies AC_{16}(n,k) \implies WO_4(n-k)} \\
\multicolumn{2}{c}{AC_1 \implies AC_{18} \implies AC_{19} \implies AC_1}
\end{array}$$

Chains such as $AC_1 \implies AC_4 \implies AC_3 \implies AC_1$ require fewer proofs than proving equivalence for every pair of definitions. We have occasionally deviated from Rubin and Rubin in order to form such chains. We have proved $AC_1 \implies AC_4$ to avoid having to prove $AC_1 \implies AC_3$ and $AC_3 \implies AC_4$. Similarly we have proved $AC_8 \implies AC_9$ instead of $AC_8 \implies AC_1$ and $AC_1 \implies AC_9$. Our new proofs are based on ideas from the text.

Creating one giant chain would minimize the number of proofs, but not necessarily the amount of effort required. In any event, we wished to avoid major deviations from Rubin and Rubin.

### 7.3. DIFFICULTIES WITH THE DEFINITIONS

Although the idea of this study was to reproduce the original proofs faithfully, we sometimes changed basic definitions in order to simplify the Isabelle proofs.

A fundamental concept is that of a *well-ordering*. The Rubins state that a set $A$ is well-ordered by a relation $R$ if $A$ is partially ordered by $R$, and every non-empty subset of $A$ has an $R$-first element; they define a partial ordering to be transitive, antisymmetric and reflexive. Isabelle/ZF defines a well-ordering to be a total ordering that is well-founded, and hence irreflexive. Fortunately there was no need to define well-ordering once again. Reflexivity does not play a major role in the Rubins' proofs, which remain valid under the Isabelle definitions. Thus, we may take advantage of the many theorems about well-ordered sets previously proved in Isabelle/ZF.

Another difference is the definition of ordinal numbers. Rubin and Rubin use essentially the same definition as Kunen does; recall §3.2. We tackle this problem by proving that their definition follows from the Isabelle/ZF one.

The Rubins use $A \prec B$ without defining it. Fortunately, its definition is standard; see §4.4 for its Isabelle formalization.

Some proofs rely on the notion of an *initial ordinal*. However, an initial ordinal is precisely a cardinal number, as previously formalized in Isabelle. After proving the appropriate equivalence we decided to use cardinals.

## 7.4. General Comments on the Proofs

We are aiming to reproduce the spirit, not the letter, of the original material. For instance, we have changed "$P(m) \implies P(m-1)$ for all $m \geq 1$" to "$P(\text{succ}(m)) \implies P(m)$ for all $m$." Such changes streamline the formalization without affecting the ideas.

Most of the implications concerning the Well-Ordering Theorem are easy to prove using Isabelle. Rubin and Rubin describe some of them as "clear." They do not prove the implication $\text{WO}_1 \implies \text{WO}_2$, but cite an external source instead. This implication is trivial with the help of Isabelle's theory of order types (recall §4.2).

It is easy to see that $\text{WO}_7$ is equivalent to the statement

If $x$ is infinite, then there exists a relation $R$ such that $R$ well-orders $x$ but $R^{-1}$ does not.

The Rubins observe (page 5) that this is equivalent to the Well-Ordering Theorem because every transfinite ordinal is well-ordered by $<$ (the membership relation) and not by $>$ (its converse). To turn this observation into a proof, we need to extend it to every well-ordered set. It is enough to prove that if a set $x$ is well-ordered by a relation $R$ and its converse, then its order type (determined by $R$) is well-ordered by $>$; this is a contradiction if $x$ is infinite. Again we exploit Isabelle order types and ordinal isomorphisms.

Rubin and Rubin's proof of $\text{AC}_7 \implies \text{AC}_6$ (page 12) fails in the case of the empty family of sets. The proof of $\text{AC}_{19} \implies \text{AC}_1$ (page 18) fails for a similar reason. When building a mechanized proof we are obliged to treat degenerate cases, however trivial they are.

The proof of $AC_9 \implies AC_1$ (page 14) has a small omission. We start with a set $s$ of non-empty sets, and define $y \stackrel{\text{def}}{=} (\cup s)^\omega$. It can be proved that for each $x \in s$, $x \times y \approx y$. Then Rubin and Rubin claim "it is easy to see that for each $x \in s$, $x \times y \approx (x \times y) \cup \{0\}$." But if $s = \{\{b\}\}$ then $x$ and $y$ are unit sets ($\{b\}$ and $\{b\}^\omega$, respectively) and the claim fails. In order to mechanize this proof we have used $x \times y \times \omega$ instead of $x \times y$. This seems simpler than handling the degenerate case separately.

On page 14, Rubin and Rubin set out to prove that $AC_{10}$ to $AC_{15}$ are equivalent to the Axiom of Choice. They describe a number of implications as "clear."[8] Then they list some implications that they are going to prove. It appears that they intend to establish two chains

$$WO_1 \implies AC_{10}(n) \implies AC_{11} \implies AC_{12} \implies AC_{15} \implies WO_6$$
$$AC_{13}(n) \implies AC_{14} \implies AC_{15}.$$

Because of other results, it only remains to show that $AC_1$ implies $AC_{13}(n)$. We could prove

$$AC_1 \implies AC_{13}(1) \qquad AC_{13}(m) \implies AC_{13}(n) \quad \text{if } m \leq n$$

or, more directly, $AC_{10}(n) \implies AC_{13}(n-1)$. In this welter of results, Rubin and Rubin have stated and we have mechanized more proofs than are strictly required.

Another noteworthy proof (page 15) concerns the implication $WO_2 \implies AC_{16}$. Rubin and Rubin devote just over half a page to it, but mechanizing it took a long time. Near the beginning of the proof they note that if $s$ is an infinite set equipollent to a cardinal number $\omega_\alpha$ then for all $k > 1$ the set of all $k$-element subsets of $s$ is also equipollent to $\omega_\alpha$. Demonstrating this is non-trivial, requiring among other things the theorem $\kappa \otimes \kappa = \kappa$ discussed above in this paper.

The next and key step is a recursive construction of a set $t = \bigcup_{\gamma < \omega_\alpha} T_\gamma$ satisfying $AC_{16}$. Now $T_\gamma$ is an increasing family of sets of $n$-element subsets of $s$. At every stage we add at most one subset. The authors claim that at any stage $\gamma < \omega_\alpha$ we can choose $n - k$ distinct elements of the set $s - (\bigcup T_\gamma \cup k_\gamma)$ where $k_\gamma$ is a $k$-element subset of $s$. They may regard this claim as obvious but we found it decidedly not so.

The difficulty of this proof lies in the complexity of the recursive definition of $T_\gamma$, which furthermore contains a typographical error.[9] Formalizing the definition was simple, but proving that it satisfied the desired property required handling theorems with many syntactically complex premises. We changed the definition several times so as to simplify these proofs.

7.5. CONSOLIDATING SOME PROOFS

Three of the Rubins' proofs, namely $AC_1 \implies WO_2$, $AC_{17} \implies AC_1$ and $AC_{15} \implies WO_6$, are based on the same idea. They construct a recursive mapping of ordinal numbers to a set. Then they show that the converse of the mapping is injective, obtaining a bijection with some desired ordinal. The mappings differ in their details. But we managed to generalize them so as to deal with only one definition, and prove some properties for one mapping instead of three.

- In the proof of $AC_1 \implies WO_2$, we start with a choice function $f$ on a set of non-empty subsets of a set $x$. We define $f(0) = u$ for some $u \notin x$ (we chose $u = x$, exploiting the Axiom of Foundation). Finally we define a mapping $G$ such that $G(\alpha) = f(x - G\text{``}\alpha)$ for all ordinals $\alpha$. (Recall that `` is the image operator).

- In the proof of $AC_{15} \implies WO_6$, we start with a function $g$ such that for every non-empty subset $y$ of a set $x$, the set $g(y)$ is a non-empty subset of $y$. We define $g(0) = u$ for some $u \not\subseteq x$ (we chose $u = \{x\}$). Then we construct a mapping $G$ such that $G(\alpha) = g(x - \bigcup_{\beta < \alpha} G(\beta))$ for all ordinals $\alpha$.

- The proof of $AC_{17} \implies AC_1$ differs from the first one in that $f$ is not necessarily a choice function, but for every non-empty subset $y$ of $x$ it satisfies $f(y) \in x$. Moreover, a mapping $H$ constructed here differs from $G$ in that it maps to $u$ every $\alpha$ such that $f(x - H\text{``}\alpha) \notin x - H\text{``}\alpha$ (which never holds if $f$ is a choice function).

For each of these definitions, Rubin and Rubin prove that the inverse of the constructed mapping $G$ (or $H$) is injective on some set, and that there is an ordinal $\alpha$ such that $G(\alpha) = u$, which somehow implies the desired result.

For the sake of clarity and economy, we decided to generalize the three definitions into one and to prove the required properties only once. Let $x$ be a set and $f$ a function such that for every non-empty subset $y$ of $x$, the set $f(y)$ is a subset of $x$. Define a mapping $H$ as follows: for every ordinal $\alpha$,

$$H(\alpha) = \begin{cases} f(z) & \text{if } f(z) \in z, \text{ where } z = x - \bigcup_{\beta < \alpha} H(\beta) \\ \{x\} & \text{otherwise} \end{cases}$$

The Isabelle definition is as follows:
```
HH(f,x,a) == transrec(a, %b r. let z = x - (UN c:b. r`c)
                                in if(f`z:Pow(z)-{0}, f`z, {x}))
```
This definition requires some adjustments to the original proofs. For $AC_1 \implies WO_2$ and $AC_{17} \implies AC_1$, the function $f$ must be replaced by a function $f'$ such that $f'(y) = \{f(y)\}$ for all $y$ in the domain of $f$. It is clear that in $AC_1 \implies WO_2$ and $AC_{15} \implies WO_6$, if $z = x - \bigcup_{\beta<\alpha} H(\beta)$ then $f(z) \in z$ holds whenever $z \neq 0$. This demonstrates agreement with the original definitions.

## 7.6. THE AXIOM OF DEPENDENT CHOICE

At the end of Chapter 2, Rubin and Rubin present two formulations of another axiom, Dependent Choice:

DC($\alpha$) If $R$ is a relation between subsets and elements of a set $X$ such that $y \prec \alpha \to \exists_{u \in X}\, y\, R\, u$ for all $y \subseteq X$ then there is a function $f \in \alpha \to X$ such that $f\text{``}\beta\, R\, f(\beta)$ for every $\beta < \alpha$.

DC If $R$ is a non-empty relation such that $\text{range}(R) \subseteq \text{dom}(R)$ then there is a function $f$ with domain $\omega$ such that $f(n)\, R\, f(n+1)$ for every $n < \omega$.

They then comment "It is easy to see that DC $\iff$ DC($\omega$)." But the only proof we could find is complicated; mechanizing it required over 200 commands. That is four times the number required for the two theorems proved explicitly.

Consider the proof of DC $\to$ DC($\omega$). Let $R \subseteq \mathcal{P}(X) \times X$ satisfy the hypothesis of DC($\omega$). Construct a set $X'$ and a relation $R'$ by[10]

$$X' = \bigcup_{n \in \omega}\{f \in n \to X \mid \forall_{k \in n}\, f\text{``}k\, R\, f(k)\}$$
$$f\, R'\, g \iff \text{dom}(g) = \text{dom}(f) + 1 \text{ and } g \restriction \text{dom}(f) = f. \qquad (f, g \in X')$$

It is easy to see that these satisfy the hypotheses of DC, which thus yields a function $f' \in \omega \to X'$ such that $f'(n)\, R'\, f'(n+1)$ for $n \in \omega$. The desired function $f \in \omega \to X$ is now defined by

$$f(n) = f'(n+1)(n).$$

A similar construction yields the converse.

The Rubins then prove, Theorem 2.20, that the Axiom of Choice (in fact, $WO_1$) implies DC($\alpha$) for every ordinal $\alpha$. While mechanizing this theorem we noticed that it is incorrect: the quantification should be restricted to cardinals. If $\alpha$ is not a cardinal then DC($\alpha$) fails.

Here is a short proof of $\neg$DC($\omega + 1$). Let $X = \omega$ and define $R$ by

$$y\, R\, u \iff y \subseteq X, y \prec \omega + 1 \text{ and } u \text{ is the least element of } X - y.$$

Assume DC($\omega + 1$). Then there is a function $f \in \omega + 1 \to \omega$ such that $f\text{``}n\, R\, f(n)$ for every $n \in \omega$; this implies $f(n) = n$. Thence $f\text{``}\omega = \omega$, so there is no $u$ such that $f\text{``}\omega\, R\, u$ as there is no $u \in \omega - \omega = \emptyset$. So DC($\omega + 1$) yields a contradiction.

## 8. Proving $\mathrm{WO}_6 \implies \mathrm{WO}_1$

The proof (page 2) of $\mathrm{WO}_6 \implies \mathrm{WO}_1$ seems to be the most complicated in the first two chapters of Rubin and Rubin. It depends upon many other properties concerning ordinal sum, equipollence, dominance, etc. To formalize some of the functions requires the description operators `LEAST` and `THE`. But the main cause of difficulty in this proof is its sheer size and complexity.

### 8.1. THE IDEA OF THE PROOF

The main idea of the proof is to show first, that every set $y$ satisfying $y \times y \subseteq y$ can be well-ordered, and then that every set $x$ can be well-ordered as a subset of such a $y$. The latter part of the proof is much easier then the former. For every set $x$ there exists a $y$ such that $x \cup (y \times y) \subseteq y$. The set $y$ is constructed as $\bigcup_{n=0}^{\infty} z_n$, where $z_0 = x$ and $z_{n+1} = z_n \cup (z_n \times z_n)$.

The main part of the proof is the claim (2), which suffices to show that every set $y$ such that $y \times y \subseteq y$ can be well-ordered:

$$\text{If } y \times y \subseteq y \text{ and } m > 1, \text{ then } m \in N_y \text{ implies } m - 1 \in N_y \qquad (2)$$

where

$$N_y = \left\{ m \mid \exists_{f,\alpha} \operatorname{dom}(f) = \alpha, \bigcup_{\beta < \alpha} f(\beta) = y, \forall_{\beta < \alpha} f(\beta) \precsim m \right\}.$$

To prove this, the Rubins assume that $y$ and $m$ satisfy the hypothesis, and that $\alpha$ and $f$ satisfy the conditions of the definition of $N_y$ for some natural number $m$. Then for every $\beta, \gamma, \delta < \alpha$ they define

$$u_{\beta\gamma\delta} \stackrel{\text{def}}{=} (f(\beta) \times f(\gamma)) \cap f(\delta).$$

It is easy to see that $\operatorname{dom}(u_{\beta\gamma\delta})$, $\operatorname{range}(u_{\beta\gamma\delta})$ and $u_{\beta\gamma\delta}$ each have no more than $m$ elements. The proof divides into two cases. For each case we construct a function $g$ satisfying the definition of $N_y$ for $m - 1$. The required ordinal number is $\alpha + \alpha$, where in this section $+$ denotes ordinal sum.

- **Case 1**: $\forall_{\beta < \alpha} . \ f(\beta) \neq 0 \to \exists_{\gamma, \delta < \alpha} . \ \operatorname{dom}(u_{\beta\gamma\delta}) \neq 0 \land \operatorname{dom}(u_{\beta\gamma\delta}) \prec m$

  Define, for $\beta < \alpha$,

  $$v_\beta = \begin{cases} \operatorname{dom}(u_{\beta\lambda_\beta\mu_\beta}) & \text{if } f(\beta) \neq 0 \\ 0 & \text{if } f(\beta) = 0 \end{cases}$$

  where $\lambda_\beta$ and $\mu_\beta$ are the lexicographically smallest pair of ordinals $\gamma$ and $\delta$ such that $\operatorname{dom}(u_{\beta\gamma\delta}) \neq 0$ and $\operatorname{dom}(u_{\beta\gamma\delta}) \prec m$.

Define the function $g$ for $\beta < \alpha + \alpha$:

$$g(\beta) = \begin{cases} v_\beta & \text{if } \beta < \alpha \\ f(\gamma) - v_\gamma & \text{if } \beta = \alpha + \gamma \end{cases}$$

Every non-empty $f(\beta)$ is split into two non-empty subsets, decreasing the number of elements.

- **Case 2**: $\exists_{\beta<\alpha} . \ f(\beta) \neq 0 \land \forall_{\gamma,\delta<\alpha} . \ \text{dom}(u_{\beta\gamma\delta}) \neq 0 \to \text{dom}(u_{\beta\gamma\delta}) \approx m$

  Let $\beta < \alpha$ be given. Since $f(\beta) \neq 0$, choose an element $s$ of $f(\beta)$. Define $v_\gamma$ for $\gamma < \alpha$ by

  $$v_\gamma = \begin{cases} \{u_{\beta\gamma\delta_\gamma}(s)\} & \text{if } f(\gamma) \neq 0 \\ 0 & \text{if } f(\gamma) = 0 \end{cases}$$

  where $\delta_\gamma$ is the smallest ordinal $\delta$ such that $\text{dom}(u_{\beta\gamma\delta}) \neq 0$. The assumptions of Case 2 justify the existence of $\delta_\gamma$ and also imply that $u_{\beta\gamma\delta_\gamma}$ is a function, justifying the notation $u_{\beta\gamma\delta_\gamma}(s)$. Only this case requires $y \times y \subseteq y$.

  Now define the function $g(\gamma)$ for $\gamma < \alpha + \alpha$ analogously to $g(\beta)$ in the previous case.

For both cases, we must show that $\alpha + \alpha$ and $g$ satisfy the definition of $N_y$ for the natural number $m - 1$. Thus we must show $\bigcup_{\beta<\alpha+\alpha} g(\beta) = y$ and $g(\beta) \precsim m - 1$ for $\beta < \alpha + \alpha$. This will complete the proof of (2).

Axiom WO$_6$ asserts that for every set $y$ there is a natural number in $N_y$. Once the claim (2) is established, it remains to apply "mathematical induction" (in fact, reverse mathematical induction) to show that $1 \in N_y$; then the function $f$ with domain $\alpha$ satisfying $f(\beta) \precsim 1$ for all $\beta < \alpha$ determines a well-ordering of $y$. Thus, if $y \times y \subseteq y$ then $y$ can be well-ordered.

8.2. PRELIMINARIES TO THE MECHANIZATION

Before mechanizing this proof, we had to prove many results in general set theory. This took a considerable time.

8.2.1. *Ordinal Arithmetic*

Both cases of the proof use ordinal sum to express $\alpha + \alpha$. At the time we conducted this proof, ordinal arithmetic was not defined in Isabelle. We adopted the following definition for ordinal sum:

```
i ++ j == ordertype(i+j, radd(i,Memrel(i),j,Memrel(j)))
```

Here `i+j` stands for disjoint sum, `radd` constructs a well-ordering on the disjoint sum (recall the discussion in §4.3) and `Memrel(i)` is the membership relation over set `i` as a set of pairs. Ordinal product is defined analogously.

The proof also makes use of ordinal difference, defined by

```
i -- j == ordertype(i-j, Memrel(i))
```

Defining the function $g$ on $\alpha + \alpha$ requires proving several properties of ordinal sum and difference. For example, if $\gamma < \alpha + \beta$ then either $\gamma < \alpha$ or $\gamma = \alpha + (\gamma - \alpha)$ and $\gamma - \alpha < \beta$. We also need the identity

```
Ord(a) ==> (UN b<a++a. C(b)) = (UN b<a. C(b) Un C(a++b))
```

These definitions of ordinal sum and product are traditional (see also Kunen [12, page 20]), but deriving the required properties from them proved to be extremely laborious. Recursive definitions [27, page 201] would have been much more direct.

### 8.2.2. *New Notation*

To express the definitions conveniently required adding a let-construct to Isabelle/ZF. Fortunately, this construct was already available in Isabelle/HOL and could be taken verbatim. A let-declaration has the syntax

```
let id = term;...; id = term in term
```

In set theory the ordering relation on ordinals coincides with the membership relation on sets: $\alpha < \beta$ means precisely $\alpha \in \beta$. But the former notation is more suggestive and most authors use it whenever possible. Isabelle has it too, with the following definition:

```
i<j  ==  i:j & Ord(j)
```

In formal proof, converting between $\alpha < \beta$ and $\alpha \in \beta$ is tiresome. We defined the quantifiers $\forall_{\beta<\alpha} . P[\beta]$, $\exists_{\beta<\alpha} . P[\beta]$ and $\bigcup_{\beta<\alpha} .P[\beta]$ for reasoning directly in terms of $<$. When $\alpha$ is indeed an ordinal, there is no difference between these and the normal bounded quantifiers.

Defining this notation took some effort; for instance, we had to ensure that the simplifier could use universally quantified assumptions as rewrite rules. It seems wrong that such trivial syntactic matters should require such effort. One might expect the proof assistant to recognize ordinals and automatically use $<$ instead of $\in$ when appropriate. However, we do not know of any system that can do this.

### 8.3. MECHANIZING THE PROOF

Figure 7 presents the Isabelle definitions of the quantities used in the proof. We used names like `NN`, `uu` and `gg1` for $N$, $u$, $g$ to avoid possible clashes with variables.

The definition of `vv1` is a formal rendering of Rubin and Rubin: "let $\lambda_\beta$ be the $<$-smallest such $\gamma$ which satisfies the conditions. Then given $\lambda_\beta$, let $\mu_\beta$ be the $<$-smallest such $\delta$ which satisfies the conditions" [24, page 3]. Unfolding the let-declarations yields nesting of the LEAST operator. To reason about this, the following theorem turned out to be useful:

```
NN(y) == m:nat. EX a. EX f. Ord(a) & domain(f)=a  &
                    (UN b<a. f'b) = y & (ALL b<a. f'b lepoll m)

uu(f, beta, gamma, delta) == (f'beta * f'gamma) Int f'delta
```

*(Case 1 definitions)*

```
vv1(f,m,b) ==
  let g = LEAST g. (EX d. Ord(d) & (domain(uu(f,b,g,d)) ~= 0 &
                            domain(uu(f,b,g,d)) lepoll m));
      d = LEAST d. domain(uu(f,b,g,d)) ~= 0 &
                   domain(uu(f,b,g,d)) lepoll m
  in  if(f'b ~= 0, domain(uu(f,b,g,d)), 0)

ww1(f,m,b) == f'b - vv1(f,m,b)

gg1(f,a,m) == lam b:a++a. if(b<a, vv1(f,m,b), ww1(f,m,b--a))
```

*(Case 2 definitions)*

```
vv2(f,b,g,s) ==
  if(f'g ~= 0, uu(f, b, g, LEAST d. uu(f,b,g,d) ~= 0)'s, 0)

ww2(f,b,g,s) == f'g - vv2(f,b,g,s)

gg2(f,a,b,s) == lam g:a++a. if(g<a, vv2(f,b,g,s),
                                    ww2(f,b,g--a,s))
```

*Figure 7.* Isabelle/ZF Definitions for $WO_6 \implies WO_1$

```
  [| P(a, b); Ord(a); Ord(b);
     Least_a = (LEAST a. EX x. Ord(x) & P(a, x))
  |] ==> P(Least_a, LEAST b. P(Least_a, b))
```

Case 2 says that "there exists an ordinal $\beta$ such that ...." The proof of this case starts with choosing the least ordinal satisfying this condition. It is not necessary for $\beta$ to be the least; any such $\beta$ can be used. This issue does not affect the informal proof. But using the LEAST operator in the formal proof would lead to needless complications.

Figure 8 presents a selection of the many lemmas that make up this proof. Most of it (theorems 1–5) involves establishing the claim (2). First, consider Case 1. Theorem 1 asserts that the union of the range of gg1 is what it should be (namely the union of the range of f). Theorem 2 asserts that each element of the range of gg1 has no more than m elements. For Case 2, theorems 4 and 5 assert analogous properties of gg2. Theorem 6 is the claim itself; theorem 7 asserts that we can construct $y$ from $x$ and theorem 8 is the final result.

Most of the definitions are made in the context of the claim (2), or its subcases. Unfortunately, in Isabelle all definitions are global. Any necessary context must be supplied explicitly. Parameters local to Case 2 of the claim

```
1  [| Ord(a);  m:nat |] ==>
   (UN b<a++a. gg1(f,a,m)`b) = (UN b<a. f`b)

2  [| Ord(a);  m:nat;
      ALL b<a. f`b ~=0 -->
      (EX g<a. EX d<a. domain(uu(f,b,g,d)) ~= 0  &
                       domain(uu(f,b,g,d)) lepoll m);
      ALL b<a. f`b lepoll succ(m);  b<a++a
   |] ==> gg1(f,a,m)`b lepoll m

3  [| ALL g<a. ALL d<a. domain(uu(f, b, g, d))~=0 -->
        domain(uu(f, b, g, d)) eqpoll succ(m);
      ALL b<a. f`b lepoll succ(m);  y*y <= y;
      (UN b<a. f`b)=y;  b<a;  g<a;  d<a;  f`b~=0;  f`g~=0;
      m:nat;  s:f`b
   |] ==> uu(f, b, g, LEAST d. uu(f,b,g,d)~=0) : f`b -> f`g

4  [| ALL g<a. ALL d<a. domain(uu(f,b,g,d)) ~= 0 -->
        domain(uu(f,b,g,d)) eqpoll succ(m);
      ALL b<a. f`b lepoll succ(m);  y*y<=y;
      (UN b<a.f`b)=y;  Ord(a);  m:nat;  s:f`b;  b<a
   |] ==> (UN g<a++a. gg2(f,a,b,s) ` g) = y

5  [| ALL g<a. ALL d<a. domain(uu(f,b,g,d)) ~= 0 -->
        domain(uu(f,b,g,d)) eqpoll succ(m);
      ALL b<a. f`b lepoll succ(m);  y*y <= y;
      (UN b<a. f`b)=y;  b<a;  s:f`b;  m:nat;  m~= 0;  g<a++a
   |] ==> gg2(f,a,b,s) ` g lepoll m

6  [| succ(m) : NN(y);  y*y <= y;  m:nat;  m~=0 |] ==> m : NN(y)

7  EX y. x Un y*y <= y

8  WO6 ==> WO1
```

*Figure 8.* Some Theorems in the Proof of $WO_6 \implies WO_1$

include $f$, $\alpha$, $\beta$ and $s$; where Rubin and Rubin write $g(\gamma)$ we must write `gg2(f,a,b,s)`g`.

In the machine proofs themselves the problem is worse. Not only the parameters, but their properties, must be copied explicitly to every lemma used in establishing Case 2. We can see this in Figure 8, theorems 3–5.

Rubin and Rubin's proof is a fine example of the difficulties of machine proof. On page 4 stand two adjacent passages, one easily mechanized, the other not. They say "Now, if in addition to $f(\beta) \neq 0$ also $f(\gamma) \neq 0$, then there exists a $\delta$ such that $u_{\beta\gamma\delta} \neq 0$. (This follows from [the definition of $u$] and the fact that $y \times y \subseteq y$.)" This statement is easily expressed in Isabelle and proved with a single call to the classical reasoner:

```
goalw thy [uu_def] "!!f. [| b<a;   g<a;   f`b~=0;   f`g~=0;
                            y*y <= y;   (UN b<a. f`b)=y
                         |] ==> EX d<a. uu(f,b,g,d) ~= 0";
by (fast_tac (AC_cs addSIs [not_emptyI]
                    addSDs [SigmaI RSN (2, subsetD)]
                    addSEs [not_emptyE]) 1);
```

They also say "if $\gamma$, $\delta < \alpha$ and $\mathrm{dom}(u_{\beta\gamma\delta}) \neq 0$ then $\mathrm{dom}(u_{\beta\gamma\delta})$ has $m$ elements. It follows ... that $u_{\beta\gamma\delta}$ has at most $m$ elements ($u_{\beta\gamma\delta} \subseteq f(\delta) \preceq m$). Therefore $u_{\beta\gamma\delta} \approx m$ and $u_{\beta\gamma\delta}$ is a function." It is obvious that if $R$ is a finite relation and $\mathrm{dom}(R) \approx R$ then $R$ is a function. But our formalization contains a long proof with numerous lemmas. The conclusion is theorem 3 of Figure 8.

Recall from §3.2 our difficulties in proving that natural numbers are cardinals. Finiteness appears to be a major source of gaps in informal proofs. When faced with an obvious statement that has no obvious proof, we are forced to prove many lemmas that look equally obvious. This is terribly frustrating. However, it appears to be a fundamental feature of formal proof, and anyway "obvious" statements are not always true!

One is reminded of the famous *mutilated chessboard* problem: if we remove two diagonally opposite corners from a chessboard, can we cover the remaining 62 squares with 31 dominos? The usual proof that the answer is "no" seems impossible to formalize without disproportionate efforts. Gardner [7] describes a number of similar puzzles.

Mechanizing the reverse induction mentioned above, and the construction from $x$ of some $y$ such that $x \cup (y \times y) \subseteq y$, is routine. All the difficulties lie in proving the claim (2). The two cases are complicated. Both authors spent considerable time experimenting with various forms of definition to make the proofs more readable.

The main file containing the proof of $\mathrm{WO}_6 \Longrightarrow \mathrm{WO}_1$ holds over 130 tactic commands; it executes in about three minutes.

## 9. Conclusions

We have mechanized parts of two advanced textbooks: most of Chapter I of Kunen [12] and the first two chapters of Rubin and Rubin [24]. Some of this material is fairly recent; the Rubins cite papers from the 1960s. In doing our proofs, we noted a number of difficulties.

Ideally, the mathematics should not have to conform to the machine: the machine should conform to the mathematics. Following a single text helps indicate whether this is indeed the case. It is not the most direct way of proceeding, however; a brief aside in the text may expand into a large formal derivation. Formalizing only the main results requires less effort while still yielding some benefits, such as finding errors and ambiguities, and exposing hidden assumptions.

On the whole, we have succeeded in reproducing the material faithfully. Isabelle's higher-order syntax makes it easy to express set-theoretic formulæ. But Rubin and Rubin frequently use English phrases that translate to complex formulæ. It is essential to ensure that the formulæ are not only correct, but as simple as possible.

Standard mathematical concepts have conflicting definitions. Sometimes these definitions are strictly equivalent, as in initial ordinals versus cardinals. Sometimes they are equivalent under certain assumptions: our definition of ordinal relies on the Axiom of Foundation. Sometimes they differ only in inessential details, as in whether a well-ordering is required to be reflexive. No details are inessential in formal proof, and we can forsee that incompatible definitions will become a serious problem as larger and larger bodies of mathematics are formalized.

Comparing the sizes of the formal and informal texts, Jutting [11, page 46] found that the ratio was constant. This may hold on average for a large piece of text, such as a chapter, but it does not hold on a line by line basis. Sometimes the text makes an intuitive observation that requires a huge effort to formalize. And sometimes it presents a detailed calculation that our tools can perform automatically. If we are going to perform such proofs on a large scale, we shall have to discover ways of predicting their size and cost.

Although set theory is formally untyped, mathematicians use different letters to range over natural numbers, cardinals, ordinals, relations and functions. There are obvious inclusions among these types: infinite cardinals are cardinals are ordinals, and all objects are sets. Isabelle's type system is of no help here. Other provers, such as IMPS [6] with its subtypes, might handle this aspect better. The proof of $WO_6 \implies WO_1$ revealed another limitation of Isabelle: its inability to allow definitions and proofs to occur within the context of a lengthy inductive argument.

Grąbczewski is engaged in proving the consistency of the Axiom of Choice, following the approach described by Kunen in Chapters 4–6. This requires coding the syntax of formulæ inductively within set theory, and internalizing the ZF axioms. Arriving at the most convenient definitions took a great deal of time. We know of no obstacle to proving deeper and deeper results in set theory, provided one is willing to devote the necessary effort.

## Acknowledgements

The research was funded by the EPSRC GR/H40570 "Combining HOL and Isabelle" and by ESPRIT Basic Research Action 6453 "Types." Grąbczewski's visit was made possible by the TEMPUS Project JEP 3340 "Computer Aided Education." John Harrison, Paul Taylor and the referees pointed out errors.

## Notes

[1] To understand those details, refer to Paulson [21, §3.5]. For $i \in I$ let $\alpha_i$ be the least $\alpha$ such that $i \in V[A]_\alpha$. From (1) we can prove

$$\frac{|I| \leq \kappa \quad I \subseteq V[A]_{\kappa^+}}{I \to V[A]_{\kappa^+} \subseteq V[A]_{\kappa^+}}$$

This result allows $V[A]_{\kappa^+}$ to serve as the bounding set for a least fixedpoint definition [19].

[2] Kunen gives straightforward inductive proofs of these first two properties. But Halmos [9, page 72] gives an argument that proves both with a single induction.

[3] Lemma 10.10 says that multiplication of finite cardinals agrees with integer multiplication.

[4] All Isabelle timings are on a Sun SPARCstation ELC.

[5] The statement of Global Choice can be obtained by Skolemizing the trivial theorem $\forall c . c \neq 0 \to (\exists x . x \in c)$. This is a standard example showing that Skolemization can be unsound in higher-order logic [13].

[6] Such figures can be regarded only as a rough guide. Many of the theorems properly belong in the main libraries. Small changes to searching commands can have a drastic effect on the run time. For comparison, the main ZF library (which includes the Kunen, Abrial and Laffitte proofs) contains 150 definitions and nearly 3300 tactic commands.

[7] Rubin and Rubin [24, page 9] state this incorrectly. They quantify over $B$ but leave $X$ free in the definiens.

[8] At least one of these, $WO_1 \implies AC_{10}(n)$, is non-trivial. We have to partition the infinite set $x$ into a set of disjoint 2-element sets, for all $x \in A$. Our proof uses the equation $\kappa = \kappa \oplus \kappa$ to establish a bijection $h$ between the disjoint sum $|x| + |x|$ and $x$. The partition contains $\{h(\text{Inl}(\alpha)), h(\text{Inr}(\alpha))\}$ for all $\alpha < |x|$.

[9] At the beginning of the fifth line from the bottom on page 15, $y \in N$ occurs instead of $y \in T_\gamma$.

[10] Here $g \restriction \text{dom}(f)$ means $g$ restricted to the domain of $f$.

## References

1. J. R. Abrial and G. Laffitte. Towards the mechanization of the proofs of some classical theorems of set theory. preprint, February 1993.
2. Grzegorz Bancerek. Countable sets and Hessenberg's theorem. *Formalized Mathematics*, 2:499–503, 1990.
   http://math.uw.bialystok.pl/~Form.Math/Vol2/dvi/card_4.dvi.
3. Robert S. Boyer and J Strother Moore. *A Computational Logic*. Academic Press, 1979.
4. Robert S. Boyer and J Strother Moore. *A Computational Logic Handbook*. Academic Press, 1988.
5. Gilles Dowek et al. The Coq proof assistant user's guide. Technical Report 154, INRIA-Rocquencourt, 1993. Version 5.8.
6. William M. Farmer, Joshua D. Guttman, and F. Javier Thayer. IMPS: An interactive mathematical proof system. *Journal of Automated Reasoning*, 11(2):213–248, 1993.
7. Martin Gardner. *The Unexpected Hanging and Other Mathematical Diversions*. University of Chicago Press, 1991.
8. M. J. C. Gordon and T. F. Melham. *Introduction to HOL: A Theorem Proving Environment for Higher Order Logic*. Cambridge University Press, 1993.
9. Paul R. Halmos. *Naive Set Theory*. Van Nostrand, 1960.
10. Gérard Huet. Residual theory in $\lambda$-calculus: A formal development. *Journal of Functional Programming*, 4(3):371–394, 1994.
11. L.S. van Benthem Jutting. *Checking Landau's "Grundlagen" in the AUTOMATH System*. PhD thesis, Eindhoven University of Technology, 1977.

12. Kenneth Kunen. *Set Theory: An Introduction to Independence Proofs*. North-Holland, 1980.
13. Dale Miller. Unification under a mixed prefix. *Journal of Symbolic Computation*, 14(4):321–358, 1992.
14. R. P. Nederpelt, J. H. Geuvers, and R. C. de Vrijer, editors. *Selected Papers on Automath*. North-Holland, 1994.
15. Philippe Noël. Experimenting with Isabelle in ZF set theory. *Journal of Automated Reasoning*, 10(1):15–58, 1993.
16. Lawrence C. Paulson. Constructing recursion operators in intuitionistic type theory. *Journal of Symbolic Computation*, 2:325–355, 1986.
17. Lawrence C. Paulson. Isabelle: The next 700 theorem provers. In P. Odifreddi, editor, *Logic and Computer Science*, pages 361–386. Academic Press, 1990.
18. Lawrence C. Paulson. Set theory for verification: I. From foundations to functions. *Journal of Automated Reasoning*, 11(3):353–389, 1993.
19. Lawrence C. Paulson. A fixedpoint approach to implementing (co)inductive definitions. In Alan Bundy, editor, *Automated Deduction — CADE-12 International Conference*, LNAI 814, pages 148–161. Springer, 1994.
20. Lawrence C. Paulson. *Isabelle: A Generic Theorem Prover*. Springer, 1994. LNCS 828.
21. Lawrence C. Paulson. Set theory for verification: II. Induction and recursion. *Journal of Automated Reasoning*, 15(2):167–215, 1995.
22. The QED manifesto. http://www.mcs.anl.gov/home/lusk/qed/manifesto.html, 1995.
23. Art Quaife. Automated deduction in von Neumann-Bernays-Gödel set theory. *Journal of Automated Reasoning*, 8(1):91–147, 1992.
24. Herman Rubin and Jean E. Rubin. *Equivalents of the Axiom of Choice, II*. North-Holland, 1985.
25. David M. Russinoff. A mechanical proof of quadratic reciprocity. *Journal of Automated Reasoning*, 8(1):3–22, 1992.
26. N. Shankar. *Metamathematics, Machines, and Gödel's Proof*. Cambridge University Press, 1994.
27. Patrick Suppes. *Axiomatic Set Theory*. Dover, 1972.
28. Yuan Yu. Computer proofs in group theory. *Journal of Automated Reasoning*, 6(3):251–286, 1990.